\documentclass[preprint,prd,tightenlines,nofootinbib,superscriptaddress]{revtex4}

\usepackage{color}
\definecolor{red}{rgb}{1,0,0}
\def\lesssim{\ \hbox{\raise 2pt \hbox{$<$} \kern -13pt
                     \lower 3pt \hbox{$\sim$}}\ }
\def\greatersim{\ \hbox{\raise 2pt \hbox{$>$} \kern -13pt
                     \lower 3pt \hbox{$\sim$}}\ }
\input epsf.tex
\def\desepsf(#1 width #2){\epsfxsize=#2 \epsfbox{#1}}

\usepackage{amsmath,bm}
\begin{document}
\preprint{\vbox{
\hbox{CERN-PH-TH/2007-027}
}  }

\title{Endpoint singularities in unintegrated parton distributions}
\author{F.\ Hautmann}
\affiliation{CERN, Physics Department, TH Division,
CH-1211 Geneva 23, Switzerland}
\affiliation{Institut f{\" u}r Theoretische Physik, 
Universit{\" a}t Regensburg,  D-93040 Germany}

\begin{abstract}
We examine the singular behavior  from the  endpoint 
region $x \to 1$ in parton distributions unintegrated 
in both longitudinal  and transverse   momenta. 
 We identify and 
regularize the singularities by  using the subtraction 
method, and 
compare this  with the cut-off regularization method. 
The counterterms for the distributions 
with subtractive regularization  are given 
in coordinate space by compact all-order  expressions 
in terms of eikonal-line operators. We carry out  an 
explicit calculation at one loop for the unintegrated quark 
distribution. 
We discuss the relation of the unintegrated parton distributions 
in subtractive regularization with the ordinary parton 
distributions. 
\end{abstract}


\maketitle

Parton distributions unintegrated in both longitudinal 
 and transverse   momenta  are used  in   QCD  to  analyze 
    hadron scattering problems with 
 multiple hard scales and to describe  infrared-sensitive  
 processes. 
 See \cite{jccrev03,jeppe06,metz} for reviews and references. 
These distributions represent less inclusive versions of the 
ordinary parton distributions. Accordingly, due to the 
lack of complete KLN~\cite{kln} cancellation, 
 they are  
 affected by singularities from the region 
$x \to 1$~\cite{jccrev03,brodsky01,collsop81}  corresponding 
to the  exclusive phase-space boundary.

In traditional  applications  these   singularities  
are handled   
by placing a cut-off  on the endpoint region.  
 The cut-off can be implemented as
 the Minkowski-space angle obtained by moving away from the 
lightcone  the eikonal line in the matrix element 
that defines the parton density, as in~\cite{korch} and 
in the method of~\cite{collsop81}, recently revisited 
 in~\cite{jiyuan,ji06}. It can also  be  implemented 
in terms of the  infrared cut-off 
associated with parton showering algorithms, as in 
Monte-Carlo event generators  that make use of unintegrated 
densities~\cite{junghgs,lonn,marchweb92}.  
But while 
  the cut-off regularization is well-suited for 
leading order calculations, it makes it difficult to go beyond  
leading accuracy. Furthermore,  with this method 
 the connection with ordinary  parton  distributions  and 
the lightcone limit  are not  so transparent.

A more systematic approach is based on 
subtractive regularization. A formulation 
 of the subtraction method, suitable  for treating 
 eikonal-line  matrix elements to all orders,   
is given in~\cite{jccfh}.  
In this approach the eikonal line attached to the 
field operator in the  original  matrix 
element  remains lightlike, but  the  
singularities are cancelled by counterterms 
provided by  certain  gauge-invariant 
 eikonal  correlators. 
The purpose of this paper  is to  study 
the unintegrated quark distribution  
using the method~\cite{jccfh}. 
To this end we analyze the structure of the endpoint singularity in 
coordinate space. We carry out  
   an explicit   calculation at one loop. This     
 allows us to identify the counterterms, and    
  provides support for an all-order operator 
formula  with 
subtractive  regularization. 
We present the analysis  for 
 the quark distribution, as this contains 
the main aspects of the  endpoint dynamics, but this    
treatment   can also  be   
given in  the  case of the gluon distribution.

The analysis is given in terms of nonlocal operators. 
 The techniques of~\cite{balbra} are used  to make contact 
with the operator product expansion  in  local operators. 
Expressing the integral of  unintegrated parton  distributions
in terms of ordinary  distributions involves in general 
 nontrivial coefficient functions, as discussed 
in \cite{jcczu} for   $ \phi^3 $ theory in dimension $d = 6$ 
and in  \cite{cchlett93} for the $x \to 0$ gluon density. 
 The subtractive-method result  that 
we find has the distinctive feature that 
 the dependence on the regularization  parameters 
 introduced by the counterterms  drops out 
 in the integrated  parton distribution.

 Let us first   recall   the basic  
  behavior near the endpoint for fixed transverse
 momentum $k_\perp$ and 
lightcone momentum fraction $x$~\cite{jccrev03}. 
The unintegrated quark distribution $f(x , k_\perp)$, 
computed in a quark target with an 
infrared regulator $\rho$~\cite{jccrev03,brodsky01}, 
has the one-loop form 
\begin{equation}
\label{strucx1_1}
f_{(1)} (x , k_\perp) =  P_R (x ,  k_\perp) - \delta (1 - x) \, 
 \delta (k_\perp ) 
 \int d x^\prime  \ d k_\perp^\prime \ 
 P_R (x^\prime ,  k_\perp^\prime) 
\hspace*{0.3 cm} , 
\end{equation}
where 
\begin{equation}
\label{strucx1}
P_R (x ,  k_\perp) \sim \alpha_s \left[ {1 \over { 
 1-x } } \ 
  { 1 \over { k_\perp^2 + \rho^2} } + \{ {\rm{regular}} 
 \;\;\; {\rm{at}}   \;\;\; x \to 1 \} \right]  
\hspace*{0.3 cm}  . 
\end{equation}
The $x \to 1$ singularity is the endpoint singularity,  
 and is present for any  $k_\perp$. 
A physical observable ${\cal O}$ is constructed 
by integrating $f$ against a test function 
$\varphi$ (specifying the final state, hard 
subprocess, etc.), which yields   
\begin{eqnarray}
\label{physobs}
 {\cal O} &=&  \int dx  \ d k_\perp \  
   f_{(1)}(x,k_\perp) \ \varphi(x,k_\perp) 
\nonumber\\
   &=&
   \int dx  \ d k_\perp \ [\varphi(x,k_\perp)
   -\varphi(1,0_\perp)]\, P_{R}(x,k_\perp)  
\hspace*{0.3 cm}  . 
\end{eqnarray}
While in the inclusive case, with $\varphi$ independent of 
$k_\perp$, the  $x \to 1$  behavior in  Eq.~(\ref{physobs}) 
 simply  corresponds  to 
 the familiar $1/(1 - x)_+$ distribution 
from real/virtual cancellation, in the general case 
 uncancelled divergences are expected from the endpoint region. 

We now proceed as follows. We 
 compute this singularity in coordinate space,  
we apply the subtractive regularization method~\cite{jccfh}, 
and then going back to 
momentum space we see the     generalization of  
Eq.~(\ref{physobs})  associated with  it.

We begin by considering  the  matrix element 
\begin{equation}
\label{coomatrel}
  {\widetilde f} ( y  ) = 
{1 \over 2} \ \sum_s \
  \langle p, s |  {\overline \psi} (y  )
  V_y^\dagger ( n ) \gamma^+ V_0 ( n ) 
 \psi ( 0  ) |
  p, s \rangle  \hspace*{0.3 cm} .  
\end{equation}
Here the $\psi$'s
 are the  two quark fields evaluated at distance 
$y = ( 0 , y^- , y_\perp  )$ with arbitrary $y^-$ and $y_\perp$,
$p$ and $s$ are the  momentum and spin of the target, 
taken to be a quark with $p^\mu = ( p^+, m^2 /(2 p^+), 0_\perp)$, 
 and $V$ is the path-ordered exponential
\begin{equation}
\label{defofV}
V_y ( n ) = {\cal P} \exp \left(  i g_s \int_0^\infty d \tau \
n^\mu A_\mu (y + \tau \ n) \right)
    \hspace*{0.3 cm} .  
\end{equation}
In Eq.~(\ref{defofV}) 
$n$ is the direction of the eikonal line and $A$ 
is the gauge field, $A_\mu = A_\mu^a t^a$, with $t^a$ 
the color generators in the fundamental representation. 
The unintegrated quark distribution  is obtained from 
 the  Fourier transform 
\begin{equation}
\label{doublefou}
    f(x, k_\perp ) =
 \int {{d y^-} \over { 2 \pi} }
 {{d^{d-2}  y_\perp }  \over { (2 \pi)^{d-2} } }
 \  e^{-i x p^+ y^- + i
 k_\perp \cdot y_\perp } \ {\widetilde f} ( y)
 \hspace*{0.3 cm} , 
\end{equation}
with $d = 4 - 2 \epsilon$ the space-time dimension. 

\begin{figure}[htb]
\vspace{65mm}
\includegraphics{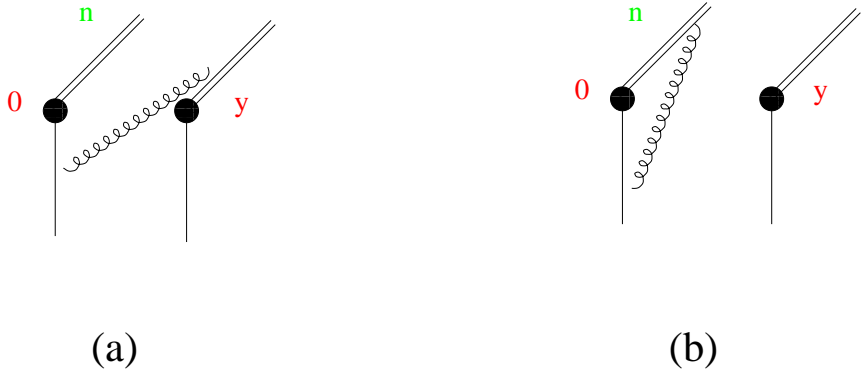}
\caption{One-loop contributions to the  quark density 
from eikonal-line couplings.}
\label{fig:w-loop}
\end{figure}

Let us  expand the  matrix element (\ref{coomatrel}) 
to one loop. In Feynman gauge, the endpoint behavior 
results from  graphs 
  with gluons  coupling to the eikonal line, 
  Fig.~\ref{fig:w-loop}(a) and (b). 
We calculate these contributions  
working  in dimensional regularization with dimension 
$d$, and 
 regulating the collinear and soft divergences
by  keeping  finite quark mass $m$ and gluon mass $\lambda$. 
  We take $n$ to be 
lightlike,  $n = (0, 1,  0_\perp )$,  as is the case 
for  ordinary parton distributions, and  will give 
later the extension of the result to $n^2 \neq 0$. 
Because we work in covariant gauge we need not consider 
the contribution from the gauge link at infinity.

 We expand the path-ordered exponential 
  in Eq.~(\ref{coomatrel}) 
to first order in $g A$, and represent the gauge-field 
two-point  correlator as 
\begin{equation}
\label{protimerep1}
\langle A_\mu^a (z_1) A_\nu^b (z_2) \rangle = 
\delta^{a b}   \ \int_0^\infty d \alpha 
 \  \int {{ d^4 \ell}  \over { (2 \pi )^4 } } \ d_{\mu \nu } ( \ell ) \ 
e^{i [ \alpha (\ell^2 - \lambda^2 + i \varepsilon ) 
+ \ell \cdot (z_2 - z_1) ]  } 
 \hspace*{0.3 cm} .   
\end{equation}
Similarly,  for quarks 
\begin{equation}
\label{protimerep2}
\langle \psi_i (z_1) {\overline \psi}_k (z_2) \rangle = 
\delta_{i k}  \ \int_0^\infty d \beta 
 \  \int {{ d^4 q}  \over { (2 \pi )^4 } } \ ({\hat q} + m) \ 
e^{i [ \beta (q^2 - m^2 + i \varepsilon) + q \cdot (z_2 - z_1) ]   } 
 \hspace*{0.3 cm} . 
\end{equation}
We switch to new 
 integration variables $v, \sigma$ 
  by setting 
\begin{equation}
\label{vandsigma}
\alpha = v \ \sigma  \hspace*{0.2 cm} , \hspace*{0.4 cm} 
\beta = (1-v) \ \sigma
 \hspace*{0.3 cm} , 
\end{equation}
with $ 0 \leq v \leq 1$, $ 0 \leq \sigma < \infty$. 
Then  the integrals in the momenta $\ell, q$ and  
$\tau , \sigma$ variables  
 can be carried out explicitly  
for the graphs in  Fig.~\ref{fig:w-loop} 
in terms of Bessel functions.    We obtain 
\begin{eqnarray}
\label{Koneloop}
 {\widetilde f}_{(a) + (b)}  &=& 
{{ \alpha_s C_F     } \over {  4^{d/2-2} \pi^{d/2-1} } } 
\ p^+ 
\int_0^1 dv  \ { v \over { 1 - v + i \varepsilon}} 
\ \left[ e^{ i p \cdot y v} \ 2^{d/2-1} \ \left( {\rho^2 \over {\mu^2
}} \right)^{d/4
-1} \ \right. 
\nonumber\\ 
&\times& \left. 
  {1 \over 
{[(-y^2 + i \varepsilon ) \mu^2 ]^{d/4-1} }} 
 K_{d/2 - 2} 
( \sqrt{\rho^2 (-y^2 + i \varepsilon )} )
- e^{ i p \cdot y } \ \Gamma ( 2 - { d \over 2} ) \ 
({ \mu^2 \over \rho^2} )^{2 - d/2}
 \right] \  ,   \hspace*{0.9 cm}
\end{eqnarray}
where  $K$ is the modified Bessel function, 
$\Gamma$ is the Euler gamma function, 
$\mu$ is the dimensional-regularization scale, 
and we have defined 
\begin{equation}
\label{rhodef}
\rho^2 = (1-v)^2 m^2 + v \lambda^2  \hspace*{0.3 cm} . 
\end{equation}
The singularity for $v \to 1$  in the integrand  
 of Eq.~(\ref{Koneloop}) 
 is the endpoint singularity. 
From the  Fourier transform (\ref{doublefou}) we see that 
$v$ has the meaning of  plus momentum fraction 
 $(p^+ - \ell^+) / p^+$, where 
 $\ell^+$ is the gluon's plus momentum.  
The $y^-$ integral from Eq.~(\ref{doublefou}) 
produces a $\delta (v - x)$
in the first term on the right hand side of 
Eq.~(\ref{Koneloop}) and a $\delta (1 - x)$ in the second
 term, thus leading to  the singularity structure for 
$x \to 1$ schematized in Eq.~(\ref{strucx1}). 
Eq.~(\ref{Koneloop}) shows in particular that the 
$v \to 1$ singularity is present even with finite  
  $\lambda$  and $m$ regulating the soft and collinear 
  regions. 

We can    see the relation  of this result with 
ordinary  parton distributions   by    expanding  
the answer (\ref{Koneloop})  in   powers of 
the distance $y^2$ from the lightcone. 
This step is analogous to the 
technique~\cite{balbra}  
to analyze nonlocal string-like operators. 
We use the   
representation of $K$ 
\begin{eqnarray}
\label{Kexpand}
&& \left( {\rho^2 \over {-y^2 + 
i \varepsilon}} \right)^{d/4
-1} \  K_{d/2 - 2} 
( \sqrt{\rho^2 (-y^2 + i \varepsilon )} )   
= 
2^{1-d/2} \ \Gamma ( 2 - { d \over 2} ) \ 
(\rho^2 )^{d/2-2} \ 
\\
&\times&
\left[ 1 + \sum_{k=1}^{\infty} b_k 
\left( - {{y^2 \rho^2} \over 4} \right)^k  \right] 
+
2^{d/2-3} \Gamma (  { d \over 2} - 2 ) \ 
(- y^2 + i \varepsilon )^{2 - d/2} \ 
\left[ 1 + \sum_{k=1}^{\infty} c_k 
\left( - {{y^2 \rho^2} \over 4} \right)^k  \right]  
  , 
\nonumber 
\end{eqnarray}
with  
\begin{equation}
\label{bkck}
b_k = { { \Gamma(  d/2 -1 ) } \over 
{ k! \ \Gamma( k + d/2 -1) }} 
\hspace*{0.2 cm}  ,  \hspace*{0.4 cm}  
c_k = { { \Gamma(  3-d/2  ) } \over 
{ k! \ \Gamma( k + 3 - d/2 ) }} \hspace*{0.4 cm}  .
\end{equation}
Inserting Eqs.~(\ref{Kexpand}),(\ref{bkck})  
in   Eq.~(\ref{Koneloop}) we get 
\begin{eqnarray}
\label{EbmEaexpand}
 {\widetilde f}_{(a) + (b)}   &=&   
{{ \alpha_s C_F     } \over {  4^{d/2-2} \pi^{d/2-1} } } 
\ p^+
\int_0^1 dv  \ { v \over { 1 - v + i \varepsilon}} \ 
\left\{ 
\left[ e^{ i p \cdot y v}  - e^{ i p \cdot y } \right] 
\ \Gamma ( 2 - { d \over 2} ) \ 
( {\mu^2 \over \rho^2} )^{2-d/2}
\right. 
\nonumber\\
&+& 
e^{ i p \cdot y v} \ 4^{d/2-2} \ 
\Gamma (  { d \over 2} - 2 ) \ 
[(- y^2 + i \varepsilon ) \mu^2]^{2 - d/2}
\nonumber\\
&+&
\sum_{k=1}^{\infty} 
{  { \Gamma (2-  { d / 2}  ) \ 
\Gamma(  d/2 -1 )}  \over { k! \ 4^k \ \Gamma( k + d/2 -1) }}  \ 
e^{ i p \cdot y v} \ ({\rho^2 \over \mu^2})^{ d/2+k-2}  
(- y^2 \mu^2 )^{k} 
\nonumber\\
&+& \left. \sum_{k=1}^{\infty} 
{  { 4^{d/2-2-k} \ \Gamma (  { d / 2} - 2 ) \ 
\Gamma(  3-d/2  )}  \over { k! \ \Gamma( k + 3- d/2 ) }}  \ 
e^{ i p \cdot y v} \ ({\rho^2 \over \mu^2})^k 
(- y^2 \mu^2  )^{2 - d/2+k} 
\right\} \hspace*{0.2 cm}  .
\end{eqnarray}
The expansion (\ref{EbmEaexpand}) separates 
long-distance  contributions  in $ \ln (\mu^2/\rho^2)$ 
and short-distance 
  contributions in  $ \ln (y^2 \mu^2)$. 
The first line of Eq.~(\ref{EbmEaexpand}) shows 
that the endpoint singularity $v \to 1$ cancels 
for ordinary parton 
distributions. 
At leading  power   the endpoint 
singularity  is  associated 
with the  coefficient function in the   second 
line of Eq.~(\ref{EbmEaexpand}).  
The higher order terms in the expansion are 
  ${\cal O}(y^2)^k$, with $k \geq 1$. 

Consider now  $n^2 \neq 0$ in the 
matrix element  (\ref{coomatrel}). In this case    
 the integrals in $\tau$ and $\sigma$ are not 
elementary, and lead to formulas in terms of parabolic 
cylinder functions. The result 
 can alternatively be 
given as the following integral representation,  
\begin{eqnarray}
\label{EbmEa}
 {\widetilde f}_{(a) + (b)}   &=&  
 {{  i\ e^{- i \pi d /4} } 
\over {  4^{d/2-3/2} \pi^{d/2-1} } } \  \alpha_s \ C_F 
\ ( \mu^2 )^{2-d/2} \ 
\int_0^1 dv  \int_0^\infty d \tau \ 
e^{ i  \tau p \cdot n (1-v) } 
\int_0^\infty d\sigma \ \sigma^{1 - d/2} 
e^{-i \sigma \rho^2  } \hspace*{0.4 cm} 
\nonumber\\ 
&\times&  
\left[   \left( 2  \ p \cdot n \ v   +  
{{\tau  n^2} \over {2 \sigma} } \right)   p^+ +  
 (1-v) \ m^2 \ n^+ 
 \right]  
\nonumber\\ 
&\times&  
  \left\{ e^{ i p \cdot y } 
\  e^{- i n^2 \tau^2 /(4 \sigma)} - e^{ i p \cdot y v} 
 \ e^{- i (y- n \tau)^2 /(4 \sigma) }  \right\}   .     
\end{eqnarray}
 This  can  be used to study the two lightcone limits 
$y^2 \to 0$ and $n^2 \to 0$. 
In Eq.~(\ref{EbmEa})  
the     behavior  of the integrand at $v \to 1$,  
resulting from the $\tau$ integration over the 
eikonal line (see   
 Eq.~(\ref{defofV})),  
 is  regularized 
by $n^2 \neq 0$. 
This is precisely the method~\cite{collsop81,jiyuan} 
to  handle  the endpoint singularity,   
 giving  a cut-off in $x$ at fixed $k_\perp$ of order  
$ 1- x \greatersim k_\perp / \sqrt{4 \eta} $, where 
$\eta = ( p\cdot n )^2 / n^2$. The parton distribution  obeys   
renormalization-group evolution 
equations in the cut-off 
parameter~\cite{collsop81,korch,korchangle} and depends on   
$\eta$,  
also after integration over  transverse momenta.

\begin{figure}[htb]
\vspace{70mm}
\includegraphics{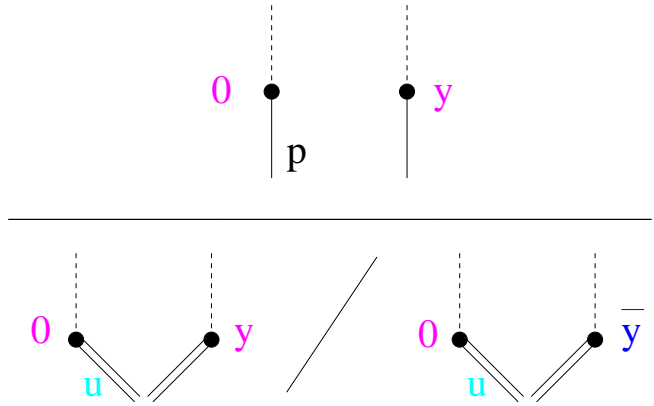}
\caption{Coordinate-space matrix element for the quark distribution 
 with subtraction factors.}
\label{fig:subtr}
\end{figure}

The subtractive method~\cite{jccfh}, which we now 
apply,  works differently.  
This is reviewed in  \cite{jccrev03}.  
The matrix element  (\ref{coomatrel}) is still evaluated at 
$n$ lightlike. It is multiplied however by vacuum 
expectation values of eikonal lines,  which 
provide counterterms for the subtraction of the endpoint 
singularity. The counterterms contain in general 
both lightlike and non-lightlike eikonals. 
For this reason 
we introduce  the vector 
\begin{equation}
\label{upum}
u^\mu = ( u^+ , u^- , 0_\perp ) 
\end{equation}
and  the path-ordered exponentials 
$V_y ( u )$, $V_{ {\bar y}} ( u )$, where $V$ is given in 
Eq.~(\ref{defofV}) and 
 $ {\bar y}$ is the lightcone projection 
 ${\bar y} = ( 0 , y^- , 0_\perp  )$.  
We consider 
the  matrix element  (Fig.~\ref{fig:subtr})
\begin{equation}
\label{defsub}
  {\widetilde f}^{\rm{(subtr)}} ( y  ) =
 {1 \over 2} \ \sum_s \
  { {
  \langle p, s |  {\overline \psi} ( y )
  V_y^\dagger ( n ) \gamma^+ V_0 ( n ) \psi ( 0 ) |
  p, s \rangle   
  } \over {
  \langle 0 |
 V_y ( u )  V_y^\dagger ( n ) V_0 ( n ) V_0^\dagger ( u ) |
  0 \rangle 
\  /  \   { \langle 0 |
 V_{ {\bar y}} ( u )   V_{ {\bar y}}^\dagger ( n )
 V_0 ( n ) V_0^\dagger ( u ) |  0 \rangle }
  }} \; ,   
\end{equation}
where $n = (0, 1,  0_\perp )$. 
The  numerator  in 
Eq.~(\ref{defsub}) 
coincides with Eq.~(\ref{coomatrel}). The denominator is the 
subtraction factor  designed to cancel 
the endpoint singularity. 
Below we verify the cancellation explicitly at one loop.
The subtraction factor  is constructed 
using the technique~\cite{jccfh}  
 and depends  on both  the lightlike 
 eikonal in direction $n$ and  the non-lightlike eikonal 
in the auxiliary  direction   $u$. 
The unintegrated quark distribution  with subtractive 
regularization is obtained as the Fourier transform 
 (\ref{doublefou}) of the matrix element (\ref{defsub}).  
This distribution depends on the direction $u$. 
However, the dependence on  $u$  cancels 
in Eq.~(\ref{defsub}) for $y_\perp =0$.

We now go back to momentum space. 
 We evaluate the Fourier transform  (\ref{doublefou})
of Eq.~(\ref{defsub}),  expanding to one loop. 
We introduce the regularization parameter 
$\zeta$, defined 
by the supplementary  eikonal in direction $u$, 
\begin{equation}
\label{zetaparam}
\zeta =  { p^{+ 2} \over 2 }  \ { u^- \over  u ^+} 
\hspace*{0.2 cm}  .
\end{equation}
At one loop the result from the matrix element 
in the numerator  is of the form in 
 Eq.~(\ref{strucx1_1}), while 
the vacuum expectation values in the denominator contribute 
subtraction terms. Explicit calculation gives 
\begin{eqnarray}
\label{subtr1loo}
\hspace*{0.5 cm}  f^{\rm{(subtr)}}_{ (1)} (x, k_\perp ) &=&
P_R (x ,  k_\perp) - \delta (1 - x) \, \delta (k_\perp )
 \int d x^\prime  d k_\perp^\prime
 P_R (x^\prime ,  k_\perp^\prime)
\nonumber\\
 &-& W_R (x , k_\perp , \zeta)
+ \delta (k_\perp )  \int d k_\perp^\prime
 W_R (x ,  k_\perp^\prime, \zeta)
\end{eqnarray}
where, restoring now also the contributions 
of finite $(1-x)$ and including 
 non-eikonal graphs, we have 
\begin{equation}
\label{prsubtr2}
P_R (x , k_\perp ) = 
{ {\alpha_s C_F} \over {  2 \pi^2 }} 
\left\{ { { (1-x) [(k_\perp^2  +
 m^2 (1 -x)^2  - 2 x m^2] }
\over { [k_\perp^2 + m^2 (1-x)^2 ]^2   }}
+ { { 2 x /{ (1-x )} }
\over {[k_\perp^2 + m^2 (1-x)^2 ]
   }} \right\}  \hspace*{0.2 cm}  , 
\end{equation}
and 
\begin{equation}
\label{prsubtr3}
    W_R (x , k_\perp , \zeta) = 
{ {\alpha_s C_F} \over {  2 \pi^2  } } 
\left\{ - { { 8 \zeta (1-x)  }
\over { [k_\perp^2 + 4 \zeta (1-x)^2 ]^2   } }
+ { { 2  /{ (1-x )}  }
\over {[k_\perp^2 + 4 \zeta (1-x)^2 ]
   }} \right\} \hspace*{0.2 cm}  . 
\end{equation}
Here we have set $\lambda = 0$, $d = 4$. 
Note the endpoint singularity 
$(1-x)^{-1} \times \alpha_s C_F /( \pi^2 k_\perp^2 )$ 
for $x \to 1$ in $P_R$, 
and the corresponding  subtraction term in  $W_R$. 
The specific form of the counterterms in the second line of 
Eq.~(\ref{subtr1loo}) 
comes from the subtraction factors in Fig.~\ref{fig:subtr}. In 
particular,  terms in $\delta (1 - x)$ 
cancel  between the  vacuum expectation values in Eq.~(\ref{defsub}). 
This reflects the fact that 
 in the 
coordinate-space results (\ref{Koneloop}),(\ref{EbmEa}) 
only the terms in $e^{ i p \cdot y v}$  
depend on $y_\perp$,  while   
 those in  $e^{ i p \cdot y}$ do not.

Observe that the unintegrated distribution 
 in Eq.~(\ref{subtr1loo})  depends 
on the parameter $\zeta$ of Eq.~(\ref{zetaparam}), but 
 upon integration in $k_\perp$
the $\zeta$-dependence cancels between 
the two terms in the second line of 
Eq.~(\ref{subtr1loo}). 
This implementation of the subtraction 
method  is to be contrasted  with
the cut-off method~\cite{collsop81,jiyuan}, where 
a residual dependence on the 
cut-off parameter is left  in the integrated distribution. 

We may now 
consider the analogue of Eq.~(\ref{physobs}) for  
  a  physical observable $\cal O$. 
Similarly to  Eq.~(\ref{physobs}),  
  suppose integrating a test function $\varphi$ 
over the distribution $f^{\rm{(subtr)}}$.  
 Using Eq.~(\ref{subtr1loo}), we obtain 
\begin{eqnarray}
\label{physobsbis}
 {\cal O} &=&  \int dx  \ d k_\perp \  
   f^{\rm{(subtr)}}_{(1)}(x,k_\perp) \ \varphi(x,k_\perp) 
\nonumber\\
 &=&
   \int dx  \ d k_\perp \ \left\{ 
( P_{R} - W_R )  \ \varphi(x,k_\perp) 
- 
  P_{R}  \ 
   \varphi(1,0_\perp)
   + W_{R}  \  \varphi(x,0_\perp)   \right\}
\nonumber\\
   &=&
   \int dx  \ d k_\perp \ \left\{ 
P_{R} \ 
[\varphi(x,0_\perp)   -\varphi(1,0_\perp)] 
+ 
(  P_{R} - W_{R} ) \ 
   [\varphi(x,k_\perp)
   -\varphi(x,0_\perp)]\,  \right\}
\hspace*{0.2 cm}  .  
\end{eqnarray}
Unlike  Eq.~(\ref{physobs}),  the endpoint behavior in 
 Eq.~(\ref{physobsbis})  is 
regularized. 
For $x \to 1$  the first term  in the last line 
 of Eq.~(\ref{physobsbis}) 
corresponds to  the 
 $1/(1 - x)_+$ distribution, 
while in the second term    the endpoint singularity in $P_R$ is  
cancelled by $W_R$.

In conclusion, we verify at one loop that 
the subtractive method, implemented in Eq.~(\ref{defsub}),  
provides a well-prescribed technique to identify 
and regularize  the endpoint singularities  in unintegrated 
 parton distributions. 
 This is accomplished 
  while keeping the eikonal in direction $n$ exactly lightlike, 
in contrast with approaches based on 
cut-off regularization~\cite{collsop81,korch,jiyuan,ji06},   
  and canceling the singularities instead  
  by multiplicative, gauge-invariant factors. 
  The one-loop counterterms  
  in Eqs.~(\ref{subtr1loo}),(\ref{physobsbis}), 
  generated from the expansion of these  
factors,    correspond  to an extension  
for 
 $k_\perp \neq 0$  of the 
plus-distribution regularization, characteristic 
of the  inclusive case. In general, 
the operator expression  in coordinate space 
 given by  Eq.~(\ref{defsub}) 
can be used to any order.

The subtractive method provides an alternative to the 
cut-off method, most commonly used in this context. 
In the cut-off  method the eikonal $n$ is moved away from the 
lightcone, as in Eq.~(\ref{EbmEa}) above, and the cut-off is 
given by $\eta = (p \cdot n)^2 / n^2$. The lightcone limit 
is not smooth, so that one does not simply recover 
the ordinary parton distribution by integrating over $k_\perp$ and 
taking $\eta \to \infty$~\cite{collsop81}. 
This behavior can be compared with the dependence on the 
``gauge-invariant cut-off" parameter 
(\ref{zetaparam}) 
 in  the case of the 
subtractive  method. As noted below Eq.~(\ref{defsub}) 
and below Eq.~(\ref{prsubtr3}), 
the dependence on the non-lightlike eikonal,  
 introduced in Eq.~(\ref{upum})  to regularize the 
 endpoint  at unintegrated level,  drops out of the 
 distribution integrated over  transverse momenta, 
 where such regularization occurs  independently  
 by KLN cancellation. 
 
Different methods of regularizing the 
endpoint singularities  in   the  unintegrated 
 parton distributions will result in different 
 coefficient functions for the $y^2$ expansion of 
 the kind  in Eq.~(\ref{EbmEaexpand}). 
 The observed  cancellation 
 of the 
 $\zeta$ dependence   in the integrated 
density  could be seen as 
 corresponding to 
 a particular  scheme choice,  which may be 
advantageous  for applications such as the construction of 
event-generation methods~\cite{jcczu,fhproc,bauermc}. It can also 
be helpful for studying the re-expansion of unintegrated 
distributions in terms of  ordinary 
distributions~\cite{balbra,jcczu,cchlett93}. 
The possibility of constructing one such scheme appears to 
be a distinctive feature of the subtractive  method  
compared to  the cut-off  method. 

\vskip 0.6 cm 

\noindent 
{\bf Acknowledgments}. I thank 
  V.~Braun, M.~Ciafaloni and J.~Collins for valuable 
  discussions.

\end{document}